# State and history in operating systems

Victor Yodaiken

Copyright 2008.*

yodaiken@finitestateresearch.com

November 23, 2018

## 1 Methods and application to operating systems

The classic UNIX code for switching processes is famously opaque and concise. In Version 6 of UNIX[5], Dennis Ritchie appended a half-hearted explanation and then added a wry: "You are not expected to understand this". Such complex state changes are at the heart of OS design. In this note, I will specify what the code does and, I hope, illustrate methods that will be of reasonably general utility in understanding and designing complex computer and software systems.

The code itself looks something like this.

```
0  switch(process_t *next){
1       if(save()){
2               resume(next);
3               panic("returned from resume");
4        } else fixmmu(); //switching in
7        return;
```

"Save" saves the process state of the current (running) process and returns "1" so that the running process then calls "resume" with a pointer to the saved process state of a second process — "next". The "resume" subroutine restores the state of the process identified by "next" and returns

"0" *as if it were returning from save.* The newly restored process then falls through to the code section marked "switching in". The saved process does not start running again until some other process calls "resume" with a pointer to its data structure.

| *Process $p_1$* | *Process $p_2$* | *Process $p_k$* |
|---|---|---|
| *save $p_1$ state* | | |
| *return 1 to "if"* | | |
| *call resume($p_2$)* | | |
| | *return 0 to "if"* | |
| | *enter else* | |
| | *fixmmu* | |
| | *return* | |
| | *...* | |
| | *switchout* | |
| | | *...* |
| | | *save $p_k$ state* |
| | | *call resume($p_1$)* |
| *return 0 to "if"* | | |

There is no time limit between a process saving and resuming and the system can get up to any number of things in-between the two operations – even suspending itself.

When process $p_1$ calls "switch" with $next = p_2$, then $p_1$ will not get to "return" until $p_2$ has resumed operation and then, in some future state, some other process $p_k$ where $p_k$ may or may not be the same as $p_2$ calls switch with $next = p_1$. One way to express this property is to say that any path $z$ that starts in process $p_1$ at the start of "switch" and that terminates in process $p_1$ at the return from "switch" must be factorable into subpaths that visit a series of intermediate states - as shown in this diagram.

```
(A, switch, enter)                           (A, switch, return)
next=B
         u1                                       u4

(A,switch,call resume)                    (C,switch, call resume)
                                                  next=A
                    u2          z      u3
                        (B,switch,fixmmu)
```

The diagram is suggestive, but it would be nice to be able to write down exactly what it means and then see how other properties interact with this property or what else we have to know about the system to assure this property.

Let $w$ be the "current event sequence" — the path that leads from the initial state of the system to the current state[1]. Let $\lambda$ be the empty sequence, $wa$ be the sequence obtained by appending event $a$ to sequence $w$, and $w \bullet z$ the sequence obtained by appending sequence $z$ to sequence $w$. Then $\lambda$ leads to the initial state. Appending an event drives the system to a successor state from the current state. Appending a sequence of events drives the system to a future state. Recursive relations are sufficient to define many event sequence dependent variables. Here's a trivial one that just counts all events.

$$Count(\lambda) = 0 \text{ and } Count(wa) = Count(w) + 1$$

In what follows, I will assume the existence of a collection of sequence dependent variables and functions that provide a window into state. Definitions of some of those functions from simpler state variables are given in section 3.

Suppose we have functions $Cline$ and $CName$ so that $Cline(p, w)$ and $CName(p.w)$ are, respectively, the current line number and current function name of process $p$ in the source code listing (assumed here to be in "C"). Then a "debugger" view of system state is given by:

$$\text{Loc}(w, p) = (CFunc(w, p), Cline(w, p))$$

For every program variable $x$, we let $V(w, p, x)$ be the current value of $x$ in the context of process $p$. For example, when $p$ is inside "switch" the value of $V(w, p, next)$ is the process identifier of the target of the switch. Note that $f(w \bullet z, g(w))$ evaluates $g$ in the state reached by $w$ and evaluates $f$ in the state reached by $w \bullet u$. So $\text{Loc}(w \bullet u, V(w, p, next))$ is the location of process $p' = V(w, p, next)$ in the state determined by $w \bullet u$.

**Proposition 1.1.** *If $\text{Loc}(w, p_1) = (switch, 0)$ and $V(w, p_1, next) = p_2 \neq p_1$ then for any $z$ so that $\text{Loc}(w \bullet z, p_1) = (switch, 7)$ there must be a process*

---

[1] There is a common theory that we have to pretend computational objects are "non-deterministic", but that seems to be based on mistaking methodological limitations for fundamental properties.

$p_k \neq p_1$ and sequences $(u_1 \bullet u_2 \bullet u_3 \bullet u_4) = z$ so that:

$$\text{Loc}(w \bullet u_1, p) = (switch, 2) \tag{1}$$
$$\text{Loc}(w \bullet u_1 \bullet u_2, p_2) = (switch, 4) \tag{2}$$
$$\text{Loc}(w \bullet u_1 \bullet u_2 \bullet u_3, p_k) = (switch, 0) \tag{3}$$
$$V(w \bullet u_1 \bullet u_2 \bullet u_3, p_k, next) = p_1 \tag{4}$$

Let's suppose we have a boolean function "Running" so that $Running(w,p) = 1$ if and only if $p$ is active.

- $Running(w,p) \in \{0,1\}$;

- $Running(w,p) > 0$ if and only if $p$ is running in the state determined by $w$;

- $Running(\lambda, p) > 0$ if and only if $p$ is running in the initial state of the system;

- $Running(w,p) > Running(wa,p)$ if and only if event $a$ causes $p$ to stop running if the system is in the state determined by $w$;

- there is a prefix $u$ of $z$ so that $Running(w \bullet u, p)$ if and only if $p$ is "sometimes" running during $z$ after the state determined by $w$.

By using event sequences we get an active view of how variables change and it is easy to define variables that help reveal the workings of a system. Here's one that counts the number of times a process has "switched in".

$In(\lambda, p) = 0,$
$In(wa, p) = \begin{cases} 1 + In(w,p) & \text{If } Running(w,p) < Running(wa,p) \\ In(w,p) & otherwise \end{cases}$

One of the advantages of the methods used here is that we are not forced to either enumerate the state set or even explain too much about the alphabet of events. For something like an OS, the event alphabet is going to be large and complex and the state set will be worse. Perhaps the event alphabet will consist of "samples" of the inputs applied to the chips of the motherboard at each processor cycle. We could imagine these events as digitized snapshots of signals. Each snapshot then indicates some discrete interval of time has passed. There may also be events that correspond to logical changes. But, for now, we can just specify the information we need to be able to decode from the event stream.

Let's require that line numbers and source code functions only change when a process is active.

$$\text{Loc}(wa, p) \neq \text{Loc}(w, p) \implies Running(w, p)$$

Note that $Running(w, p) = 1$ may not mean $Running(w, p') = 0$ because we leave open the possibility of multiple processor cores. More on that below.

Note that $V(wa, p, x) \neq V(w, p, x)$ does not necessarily imply that $Running(w, p) = 1$ — because many of the objects within the address space of a process are shared objects. For example the pages may page in or out, data may arrive from a DMA device, there may be notification of an I/O or other event, and shared data structures will be modified by other processes. Modularity in operating systems is a tough engineering challenge.

## 2 Instrumenting the OS

Proposition 1.1 is a "safety" property — it requires that *if* there is a path from entry to exit, the path must have certain properties. We also need a liveness property — that processes will advance from switch to the running of the target process.

If each event defines signals over a specified unit of time, then we can have $Time(w)$ provide the current time in some sufficiently fine unit. Without going into to much detail, $Time$ needs to behave sensibly:

$$Time(w) \leq Time(w \bullet u)$$

We will often need to count how much time passes during an event or sequence of events
$$Time(w \bullet u) - Time(w)$$
tells us how much time passes during "u" after "w" and

$$Time(wa) - Time(w)$$

measures the time during the single event $a$. It may be that there are events that take no real-time or maybe each event corresponds to a sample of signals during a discrete interval or even that event duration depends on history. We don't have to worry about any of that yet.

Let's also suppose we have $ValidProcess(w, p)$ to tell us whether a process identifier $p$ identifies an actual, instantiated process (on any core) and we have $Ready(w, p) \in \{0, 1\}$ to tell us is a process is ready to run.

$$ValidProcess(w, p) \in \{0, 1\}$$
$$Ready(w, p) \in \{0, 1\}$$
$$Running(w, p) \leq ValidProcess(w, p)$$
$$Ready(w, p) \leq ValidProcess(w, p)$$

We can now define how long a process has been waiting to run.

$$Waiting(\lambda, p) = 0$$
$$Waiting(wa, p)$$
$$= \begin{cases} (Time(wa) - Time(w)) + Waiting(w, p) & \text{if } Running(w, p) < Ready(w, p) \\ 0 & \text{otherwise} \end{cases}$$

A a system is $t_{live}$ live if $Waiting(w, p) < t_{live}$ for all $w$. Although some researchers have decided that "liveness" should be considered a property "in the limit" (without an explicit time bound), I don't think such a version of liveness means anything interesting when we are discussing engineered discrete state objects.

**Proposition 2.1.** *Calling switch forces process "next" to run within a fixed time.*
*There is a $t_{switch}$ so that for any $w$ and $z$:*

*If $\mathrm{Loc}(w, p) = (switch, 0)$ and $Time(w \bullet z) \geq Time(w) + t_{switch}$*
*then there is a prefix $u$ of $z$ so that $\mathrm{Loc}(w \bullet u, V(w, p, next)) = (switch, 5)$*

Proposition 2.1 has to be true if the system is $t_{live}$ live. Otherwise, the switching out process could stall, forever.

The two propositions formalize what we want the switch code to do at a high level, but do not specify how state must be preserved over a switch. Since process state consists of both shared and non-shared data, we have to distinguish those:

**Proposition 2.2.**

*If* $\text{Loc}(w, p) = (switch, 3)$ *and* $\text{Loc}(w \bullet u, p) = (switch, 5)$
*and there is no proper prefix $z$ of $u$ so that* $\text{Loc}(w \bullet z, p) = (switch, 5)$
*then for any non-shared variable* $x, V(w, p, x) = V(w \bullet u, p, x)$

# 3 Digging down

Here's a list of functions "assumed" into existence above that need to be either justified or defined from simpler elements.

$$Cline$$
$$Cname$$
$$SavedRegisters$$
$$StackContents$$
$$ValidProcess$$
$$Ready$$
$$Running$$
$$Time$$
$$NonShared$$
$$V$$

Let's suppose that the machine has 1 or more cores and that

$$\text{Reg}(w, c, r), \text{Mem}(w, c, loc)$$

are, respectively the contents of register $r$ on core $c$ and the contents of memory location $loc$ on core $c$. For example $\text{Reg}(w, c, PC)$ (program counter) and $\text{Reg}(w, c, SP)$ (stack pointer) are useful to know. Given a program listing $L$ and the current program counter, it is reasonably straightforward to compute *CLine* and *CName*, so I won't dig into those further. Given these values, whether a symbol is a stack or global variable is also straightforward, so we assume *IsStack* and *IsGlobal* can be constructed. Furthermore, for global variables the correspondence between name and address is determined by the program listing and some data about the compiler/linker settings. Suppose there is a memory location $current[c]$ for each core $c$ that holds the identity of the current process on core $c$. Then $\text{Mem}(w, c, current[c])$ is the process running on core $c$. We have to require that

$$\text{Mem}(w, c, current[c]) = \text{Mem}(w, c', current[c']) \leftrightarrow c = c'$$

and then

$$Running(w,p) \begin{cases} 1 & \text{if for some } c, \text{Mem}(w,c,current[c]) = p \\ 0 & \text{otherwise.} \end{cases}$$

$$Ready(w,p) \begin{cases} 1 & \text{if for any } c, Bitset(\text{Mem}(w,p+procstatus), READY) \\ & \text{and } ValidProcess(w,p) \\ 0 & \text{otherwise.} \end{cases}$$

$$V(wa,p,x) \begin{cases} \text{Mem}(wa,c,y) & \text{if Mem}(w,c,current[c]) = p \\ & \text{and } IsGlobal(w,p,x) \text{ and } y = x \\ & \text{and } IsStack(w,p.x) \text{ and } y = x + \text{Reg}(w,c,SP) \\ V(w,p,x) & \text{otherwise.} \end{cases}$$

If $\text{Reg}(w,c,SP)$ is the contents of the stack pointer register on core $c$, then $\text{Mem}(w,c,\text{Reg}(w,c,SP))$ is the contents of the top of the stack on processor core $c$ (assuming alignment and so on). In many operating systems, the kernel stack of a process, which is what we are discussing here, is fixed size and "grows down" by subtraction from a, for example, 8K boundary. One of the reasons for doing this is that its easy to calculate the stack base by $bitwiseand(stackaddress + 8095, bitinvert(8095))$ if the stack is 8K and on an 8K boundary. In that case, we can define StackContents so it captures the stack.

$$StackContents(wa,p)$$
$$= \begin{cases} (\text{Mem}(w,c,a)...\text{Mem}(w,c,b)) & \text{if } Running(w,p) \\ & \text{and } \text{Mem}(w,c,current[c]) = p \\ & \text{and } a = \text{Reg}(w,c,SP) \\ & \text{and } b = bitiseand(a+8195), bitinvert(8195)) \\ & \text{and increments between } a \text{ and } b \text{ are by } wordsize \\ StackSize(w,p) & \text{otherwise} \end{cases}$$

Note that StackContents is defined so that it does not change when the process is not running. If we dig down to the assembler level, we'd probably want to be sure that the stack contents at the point of return from save was the same as that at the point of return from resume.

| | |
|---|---|
| $Cline$ | from $\text{Reg}(w, c, PC)$ |
| $Cname$ | from $\text{Reg}(w, c, PC)$ |
| $SavedRegisters$ | from $\text{Reg}(w, c, ..)$ |
| $StackContents$ | from $\text{Reg}(w, c, SP)$ and $\text{Mem}(w, c, ..)$ |
| $ValidProcess$ | from $\text{Mem}(w, c, p->status)$ |
| $Ready$ | from $\text{Mem}(w, c, p->status)$ |
| $Running$ | from $\text{Reg}(w, c, current)$ |
| $Time$ | primitive |
| $NonShared$ | from symbol table |
| $V$ | from $\text{Mem}(w, c, ...)$ |

## 4  Parallelism and encapsulation

Parallelism is a huge issue in "formal methods" but appears naturally here. For example, it is certainly possible that for some $w$ and $a$ there are several cores $c$ so that $\text{Reg}(wa, c, PC) \neq \text{Reg}(w, c, PC)$. We have not had to yet specify anything about the way the cores change state in parallel — they just are specified in a way that makes it possible. In some cases, however, we want to describe systems in which the architecture of components is specified and that is also straightforward.

Consider an abstract model of process interaction where processes can either wait for or generate events and, only one process can advance per core. We are going to want to connect up a collection of these processes so that they communicate sychronously.

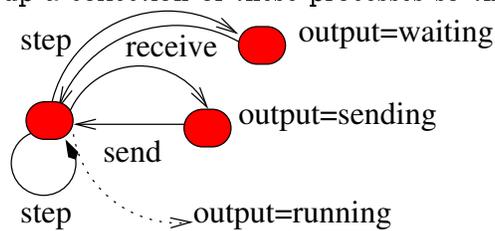

Note that the diagram obscures the intent that there may be many different states where output is running, waiting, or sending.

**Definition 4.1.** *$f$ is an abstract state process over $P$ and $X$ with id $p_0$*

*if and only if*

$$f(w) \in \{running, waiting[p], sending[x, p] : x \in X, p \in P\}$$
$$\text{and } f(w) \neq running \implies f(w \cdot \langle step \rangle) = f(w)$$
$$\text{and } f(w) = sending[x, p] \implies f(w \cdot \langle send \rangle) = idle$$
$$\text{and } f(w) = waiting[p] \implies f(w \cdot \langle receive[x, p] \rangle) = idle$$
$$\text{and } f(w) \neq waiting[p_0] \text{ — never wait for self}$$
$$\text{and } f(w) \neq sending[x, p_0] \text{ — never sent to self}$$

Many distinct sequence dependent functions can satisfy this specification. That is, we can have $A_1$ and $A_2$ that are both abstract processes by this definition where $A_1(w) \neq A_2(w)$ for some or even most $w$. An abstract process that is "running" has some internal procedure for deciding when to request to send or receive a message. We do not need, now, to decide what that process is, but it could easily be the execution of a program it receives as a message or something fixed in its internal operation or some combination. Finally, we have not specified what happens when unwanted events happen — such as a receive from $p'$ when the process wants to receive from $p$.

Now let's define a connected system of such abstract processes. Suppose that each of $A_{p_1} \ldots A_{p_k}$ are abstract processes and define

$$F(w, p) = A_p(w_p)$$

where we will define $w_p$ recursively.

$$\lambda_p = \lambda \text{ and } (wa)_p = w_p \bullet g(w, a, p)$$

and

$$g(w, a, p) = \begin{cases} \langle receive[x, q] \rangle & \text{if } A_p(w_p) = waiting[q] \\ & \text{and } A_q(w_q) = sending[x, p] \\ \langle send \rangle & \text{if } A(w_p) = sending[x, q] \\ & \text{and } A(w_q) = waiting[p] \\ \langle step \rangle & \text{if } A(w_p) = running \\ & \text{and } Running(w, p) \\ \lambda & \text{otherwise.} \end{cases}$$

Note that $p$ only gets to "step" if it is selected as the running process in the encompassing environment of the operating system.

# 5   Conclusion and mathematical note

In brief, sequence functions are representations of Moore type state machines. Given a sequence function $f$ over alphabet $B$ let $B^*$ be the set of finite sequences over $B$ including $\lambda$ and define

$$w \sim_f z \iff \forall u \in B^*, f(w \bullet u) = f(z \bullet u)$$

Then define $[w]_f = \{z : z \sim_f w\}$ and consider the set of these equivalence classes $S_f = \{[w]_f : w \in B^*\}$. Define $\delta_f([w]_f, a) = [wa]_f$ and define $\gamma_f([w]_f) = f(w)$. Then $M_f = (B, S_f, [\lambda]_f, \delta_f, \gamma_f)$ is a classical (although not necessarily finite) Moore machine with state set $S_f$, initial state $[\lambda]_f$, transition map $\delta_f$, and output map $\gamma_f$.

Conversely, given a Moore machine $M = (B, s_0, \delta, \gamma)$ define $f_M$ so that $f_M(w) = \gamma(\delta^*(w))$ where $\delta^*(\lambda) = s_0$ and $\delta^*(wa) = \delta(\delta^*(w), a)$.

The encapsulation of section 4 corresponds to a Moore machine produce called the *general product*[2]. For simplicity let's define this product for finite numbers of state machines. Suppose $f : B^* \times X \to Y$ where $X = \{x_1, \ldots x_k\}$ is defined by $f(w, x) = g(w_x)$ where $\lambda_x = \lambda$ and $(wa)_x = w_x \bullet \rho(f(w, x_1) \ldots, f(w, x_k), a, x)$. For even more simplicity, suppose $\rho(y_1, \ldots y_k, a, x_i) \in B_i$. Then for each $i$ we can construct a $M_{g_i} = (S_i, s_{0_i}, \delta_i, \gamma_i)$ using the construction above. Define a product by $M_f = (\Pi_i B_i, (s_{0_0} \ldots s_{0_k}), \delta, \gamma)$. Each state of $M_f$ is a $k$-tuple $s = (s_1, \ldots s_k) \in \Pi_i S_i$. The transition function $\delta$ is constructed as follows:

$$\delta(s, a) = (\delta_1(s_1, \rho(\gamma_1(s_1) \ldots \gamma_k(s_k), a, x_1)), \ldots \delta_k(s_k, \rho(\gamma_1(s_1) \ldots \gamma_k(s_k), a, x_k))).$$

Finally: $\gamma((s_1, \ldots s_k)) = (\gamma_1(s_1), \ldots \gamma_k(s_k))$. Then $f_{M_f}(w) = (f(w, x_1) \ldots f(w, x_k))$. It may be seen why the functional representation is advantageous in some situations.

Consideration of the algebraic basis of state machine theory and the relationship between state machines and semigroups indicates that there may be some value in looking at the algebraic structure of sequence dependent functions. If $\cong_f$ is defined so that

$$w \cong_f u \iff \forall z_1, z_2, f(z_1 \bullet w \bullet z_2) = f(z_1 \bullet u \bullet z_2)$$

then the congruence classes $[[w]]_f = \{u : w \cong_f w\}$ form a monoid under the operation $[[w]]_f \times [[u]]_f = [[w \cong u]]_f$. If we constrain $\rho$ to not depend on any feedback, so that transitions to $M_i$ depend only on outputs of $M_j : j < i$, then the results of Krohn-Rhodes theory as described in Holcombe [4], Arbib [1] and Ginzburg [3]. What happens if $\rho$ is constrained in other ways, such as by a certain circuit design discipline? Also, in databases, using some circuit disciplines, and in other situations, invertibility is a useful property. That invertibility produces sequence functions that correspond to groups.

A much earlier version of this work can be found in [9] and [8] and much earlier in [7] with applications in [6] and [10]. Unfortunately, it took me many years to understand good advice from Professor George Avrunin that the formal logic notation was an impediment instead of an advantage.